\documentclass[twocolumn,twocolappendix,tighten,times]{aastex63} 

\usepackage{tikz}
\usepackage{amsmath}
\usepackage{multirow}
\usepackage{nicefrac}
\usepackage{enumitem}
\usepackage{pbox}
\usepackage{soul}
\usepackage{xfrac}
\usepackage{float}

\let\OLDthebibliography\thebibliography
\renewcommand\thebibliography[1]{
  \OLDthebibliography{#1}
  \setlength{\parskip}{-1pt}
  \setlength{\itemsep}{-1pt plus 0.3ex}
}

\shorttitle{Water on Hot Rocky Exoplanets}
\shortauthors{Kite \& Schaefer}

\begin{document}


\title{\Large{\textbf{Water on Hot Rocky Exoplanets}}}%


\author{Edwin S. Kite}
\affil{Department of the Geophysical Sciences, University of Chicago, Chicago, IL (kite@uchicago.edu).}

\author{Laura Schaefer}
\affil{School of Earth Sciences, Stanford University, Palo Alto, CA.}


\begin{abstract}
\noindent Data suggest that most rocky exoplanets with orbital period $p$~$<$~100~d (``hot'' rocky exoplanets) formed as gas-rich sub-Neptunes that subsequently lost most of their envelopes, but whether these rocky exoplanets still have atmospheres is unknown. We identify a pathway by which 1-1.7 $R_\Earth$ (1-10 $M_\Earth$) rocky exoplanets with orbital periods of 10-100~days can acquire long-lived 10-2000 bar atmospheres that are H$_2$O-dominated, with mean molecular weight $>$10. These atmospheres form during the planets' evolution from sub-Neptunes into rocky exoplanets. H$_2$O that is made by reduction of iron oxides in the silicate magma is highly soluble in the magma, forming a dissolved reservoir that is protected from loss so long as the H$_2$-dominated atmosphere persists. The large size of the dissolved reservoir buffers the H$_2$O atmosphere against loss after the H$_2$ has dispersed. Within our model, a long-lived, water-dominated atmosphere is a common outcome for efficient interaction between a nebula-derived atmosphere (peak atmosphere mass fraction 0.1-0.6 wt\%) and oxidized magma ($>$5 wt\% FeO), followed by atmospheric loss. This idea predicts that most rocky planets that have orbital periods of 10-100 days and that have radii within 0.1-0.2 $R_\Earth$ of the lower edge of the radius valley still retain H$_2$O atmospheres. This prediction is imminently testable with JWST and has implications for the interpretation of data for transiting super-Earths.\\

\vspace{-0.05in}
\noindent \emph{Unified Astronomy Thesaurus concepts:} Extrasolar rocky planets (511); Exoplanet atmospheres (487); Exoplanet evolution (491)

\vspace{0.1in}

 \end{abstract}

\section{Introduction.}
\noindent Most sunlike stars are orbited by a hot rocky exoplanet -- a~world with orbital period $p$ $<$ 100 days, radius $R$~$<$1.7~$R_\Earth$, and a density indicating that it is mostly composed of \mbox{Fe-metal} and silicates \citep{Dai2019,Hsu2019,Otegi2020}. The question ``do these worlds have atmospheres?'' defines the observational frontier for rocky-exoplanet research (e.g., \citealt{Kreidberg2019,Koll2019}). For example, searches for high-molecular-weight atmospheres on hot rocky exoplanets will be possible using James Webb Space Telescope (JWST). Data suggest that $\sim$80\% of hot super-Earths formed as sub-Neptunes, with thick nebula-derived atmospheres of H$_2$ overlying silicate magma, and that the H$_2$ envelope (atmosphere mass fraction $f_{atm}$ up to 1 wt\% of planet mass) was removed by atmosphere loss to space (e.g. \citealt{RogersOwen2020}). Here we define a ``super-Earth'' as a world with 1~$R_\Earth$~$<$~$R$~$<$~1.7~$R_\Earth$, and a ``sub-Neptune'' as a world with 1.8~$R_\Earth$~$<$~$R$~$<$~4~$R_\Earth$. The apparent near-ubiquity of an atmospheric loss process powerful enough to halve planet volume suggests that super-Earths that formed as sub-Neptunes would not have a nebula-derived atmosphere today. In other words, the prevailing view is that observing a rocky super-Earth with a thin ($<$500~km thick) nebula-derived atmosphere would require fine-tuning, either of the time of cessation of the atmosphere loss process, or of the time of observation relative to the planet's evolution. 

Thus, previously proposed routes by which a hot rocky super-Earth might acquire an atmosphere have focused on solid-derived volatiles (involving volatile-rich bolides, e.g. \citealt{Bitsch2019}, or volcanism from a volatile-bearing interior, e.g. \citealt{KiteBarnett2020}). Whether or not these mechanisms robustly lead to atmospheres on hot rocky exoplanets depends on the details of volatile transport from colder regions of the exoplanetary system. These mechanisms do not require an initial atmosphere of nebula-derived H$_2$. These mechanisms would work as well or better for planets that formed without such an atmosphere (``intrinsically rocky'' worlds). 

By contrast, in this study we explore a pathway which only endows rocky planets with atmospheres if those planets are born with thick H$_2$ envelopes.\footnote{In this study, we use the terms ``atmosphere'' and ``envelope'' as synonyms.}
Specifically, we investigate a pathway by which atmospheric escape distills the products of chemical reactions between silicate magma and nebula-derived H$_2$ to yield hot rocky exoplanets with 10-2000~bar, long-lived H$_2$O atmospheres. Surprisingly, as we will show, this distillation happens even when the atmospheric escape is itself unfractionating. 

\vspace{-0.00in}
\section{Method.}

\noindent Our method shares the assumptions of \citet{Kite2020}, with the major change being the incorporation of atmospheric escape in the present study. \citet{Kite2020} calculate magma-atmosphere equilibration in the Fe-Mg-Si-O-H system, neglecting Fe$^{3+}$ and He, and assuming that volatiles equilibrate with magma at $T$~$\sim$~2500~K. H$_2$ from the nebula is oxidized by magmatic Fe$^{2+}$O to form H$_2$O (this has previously been proposed as a way to form oceans on habitable-zone planets, e.g. \citealt{Sasaki1990, IkomaGenda2006}). The key reaction is 

\vspace{-0.075in}
\begin{equation}
\mathrm{FeO_{(l)} + H_{2(g)} = Fe_{(l)} + H_{2}O_{(g)}}
\end{equation}

\noindent Following \citet{Kite2020}, we assume here that the initial nebula-derived atmosphere chemically equilibrates with a well-stirred magma ocean at the pressure and temperature of the magma-atmosphere interface. Using the wide range of pre-volatile-equilibration magma FeO weight fractions that is plausible for exoplanets \citep{ElkinsTantonSeager2008,Doyle2020}, the resulting H$_2$/H$_2$O atmosphere has a mean molecular weight ($\mu_{atm}$) of 2~-~7~Da \citep{Kite2020}. Fe$^{2+}$~reduction yields liquid metal (Fe$^0$), which is insoluble in magma, and sinks to the planet's center where (we assume) it is chemically isolated. Both H$_2$O  and H$_2$ are soluble in magma, but there is a big contrast in the solubilities. Taking basaltic magma at 1 GPa as an example, in the presence of pure H$_2$O, H$_2$O solubility is  16~wt\% (extrapolating \citealt{Papale1997}), but in the presence of pure H$_2$, H$_2$ solubility is only 0.57 wt\%  \citep{Hirschmann2012}.

A large fraction of the planet's total H$_2$ is stored in the atmosphere, but almost all of the H$_2$O (10$^{23}$-10$^{24}$ kg, equivalent to 10$^2$-10$^3$ Earth oceans) is stored in the magma. This is much larger than the mass of other volatile species, such as C, which we neglect.

In this study we ask: \emph{Under atmospheric loss, does the H$_2$O escape to space, re-oxidize the magma, or stay as a H$_2$O envelope?} To answer this question, we first model atmospheric loss from the model planets calculated by \citet{Kite2020}. (Loss could be via hydrodynamic escape, itself powered either by X-ray and Extreme UltraViolet radiation from the star or by the luminosity of the silicate core, or via impact erosion; \citealt{Bean2020}). We model bulk (nonfractionating) loss, as the atmosphere is too hot for water to condense or to unmix from H$_2$, and the loss rate is thought to be too fast for diffusive seperation of H$_2$ from H$_2$O to be important (e.g., \citealt{Hu2015}, \citealt{KiteBarnett2020}). (We return to the role of selective escape in \S4). 

We carry out bulk removal of atmosphere from the planet via many small steps. After each small step of gas removal, we re-solve for atmosphere-magma equilibrium partitioning of each of the two volatiles (H$_2$ and H$_2$O):

\vspace{-0.15in}
\begin{align}
\begin{split}
c_i &= e_i + p_i s_i m_{magma} \\
      &= e_i + [(e_i / A_{mai}) g (\mu_{atm}/\mu_i)]s_i m_{magma} 
      \end{split}
      \label{eqn:partition}
\end{align}

\noindent Here, $c_i$ is the total inventory of the volatile $i$ (decreasing at each step due to removal to space), $e_i$ is the mass in the
atmosphere, $p_i$ is the partial pressure at the magma-atmosphere interface, and $A_{mai}$ (and $g$) are the area of (and gravitational acceleration at) the magma-atmosphere interface. $\mu_{atm}$
is the mean molecular weight of the atmosphere, $s_i$ is the solubility coefficient
(mass fraction Pa$^{-1}$) of the volatile, and
$m_{magma}$ is the magma-ocean mass.
Each step depressurizes the top of the magma ocean. Therefore, at each step some gas in the ocean is exsolved. Exsolution is a brake on thinning of the atmosphere ($\partial e_i / \partial c_i < 1$). This ``brake'' is more effective for H$_2$O than for H$_2$ because of the large mass of the dissolved-in-magma H$_2$O reservoir. 

\begin{figure}
\centering
\includegraphics[width=1.0\columnwidth,clip=true,trim={0 0 0 0}]{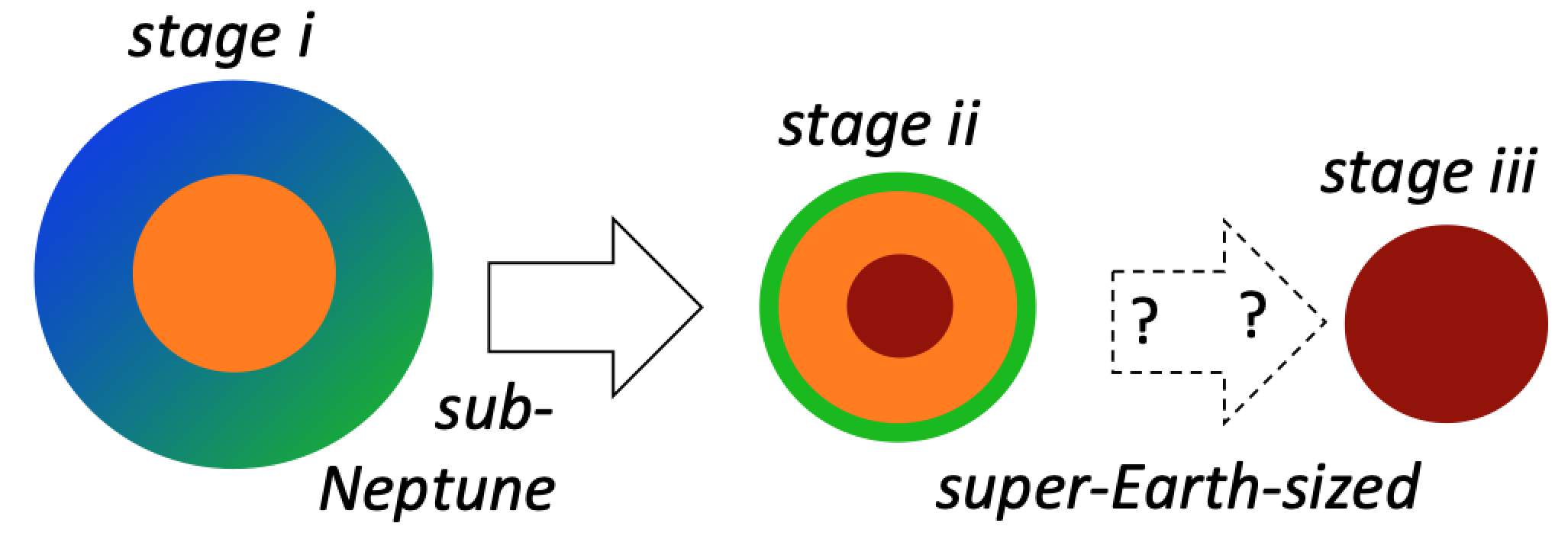}
\includegraphics[width=1.02\columnwidth,clip=true,trim={3 0 3 4}]{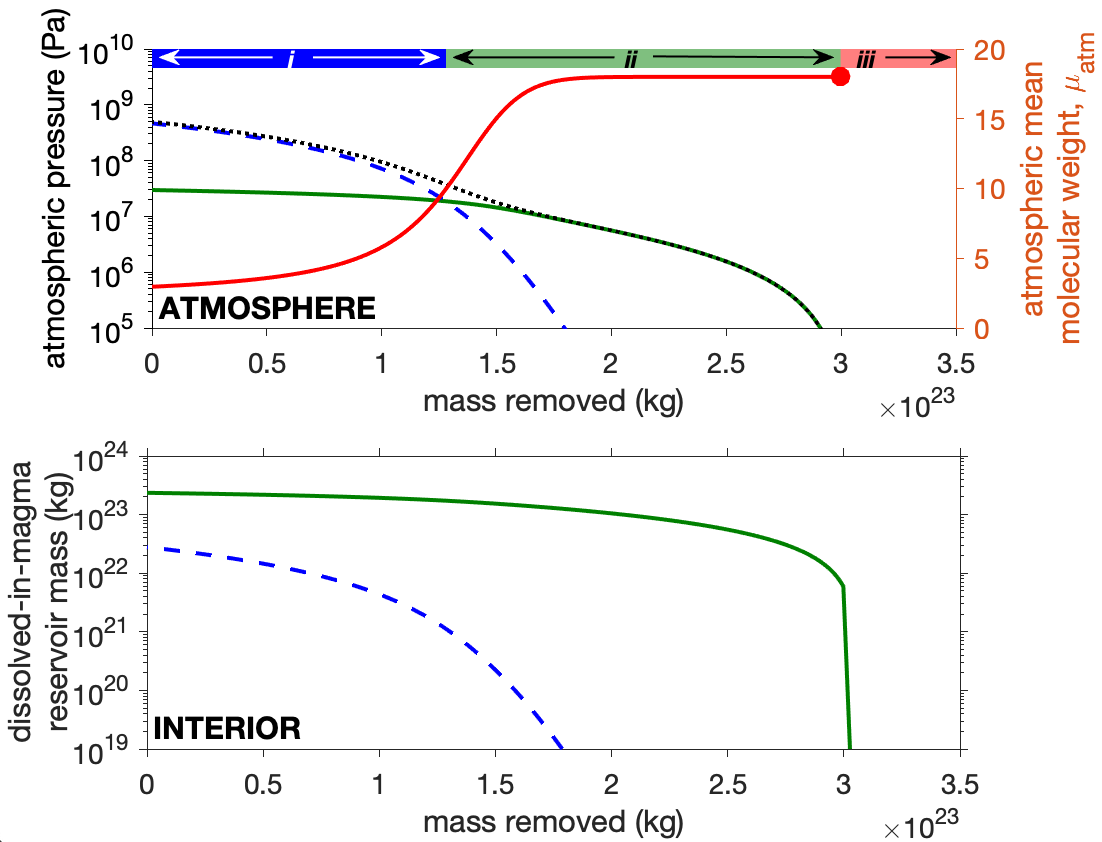}
\caption{Example model output. \emph{Top row:} Overview of the modeled scenario: a sub-Neptune with a core of mass 5~$M_{\Earth}$ that initially has FeO content similar to that of Earth's mantle, then has H$_2$ added until the base-of-atmosphere-pressure is 5 kbar (with magma-atmosphere equilibration) (stage i). Blue and green shading corresponds to atmospheric H$_2$ and atmospheric H$_2$O, respectively. The planet then undergoes non-fractionating (bulk) atmospheric loss and the atmosphere evolves into a H$_2$O atmosphere (stage ii). Orange shading corresponds to liquid magma and red shading corresponds to solid rock. The Fe-metal core is not shown. \emph{Lower panels:} The corresponding H$_2$ reservoirs (dashed blue lines), H$_2$O reservoirs (green lines), total atmospheric pressure (dotted line) and atmospheric mean molecular weight (red line) for this scenario. The blue bar (``i'') corresponds to the H$_2$-dominated regime, the green bar (``ii'') corresponds to the H$_2$O-dominated regime, and the red bar (``iii'') corresponds to the bare-rock stage after all atmosphere has been lost. The H$_2$O-atmosphere stage (`ii'') is long because the H$_2$O atmosphere is buffered by the large dissolved-in-magma H$_2$O reservoir.
}
\label{fig:fig1}
\end{figure}

In general $s_i$ varies with $p_i$ and total pressure. For this study, we follow \citet{Schaefer2016} and set $s_{\mathrm{H2O}}$~=~3.44~$\times$~10$^{-8}$~$p_{\mathrm{H2O}}^{-0.26}$. We assume that   $s_{\mathrm{H2}}$ is equal throughout each simulation to its value at the maximum atmospheric pressure for each simulation. This overstates the shielding of H$_2$ within the magma. This is conservative relative to our conclusion that H$_2$O-dominated atmospheres emerge relatively early.

\section{Results}
\noindent Fig.~\ref{fig:fig1} shows results for a world with mass $M$~=~5~$M_{\Earth}$ that initially has magma $X_{\mathrm{FeO}}$~= 11~wt\%, similar to the FeO content of Earth's mantle. In order to~generate the initial condition (left side of Fig.~\ref{fig:fig1}), within our model H$_2$ is added to the model planet, and magma-atmosphere interaction occurs, forming H$_2$O. The initial H$_2$ mass is specified such that the model atmospheric pressure (following magma-atmosphere interaction) is 5 kbar. The resulting model sub-Neptune has $\mu_{atm}$ $\approx$ 3, $X_{\mathrm{FeO}}$~=~6~wt\%, and a radius of roughly $\sim$1.9~$R_\Earth$ \citep{LopezFortney2014}. $\sim$1.9~$R_\Earth$ worlds are underabundant in the exoplanet census, and there is strong statistical evidence that this is due to atmospheric loss, forming $R$~$<$1.7~$R_\Earth$ planets (e.g., \citealt{RogersOwen2020}). 

\begin{figure}
a)
\includegraphics[trim={0mm 0mm 0mm 0mm},clip,width=1.0\columnwidth]{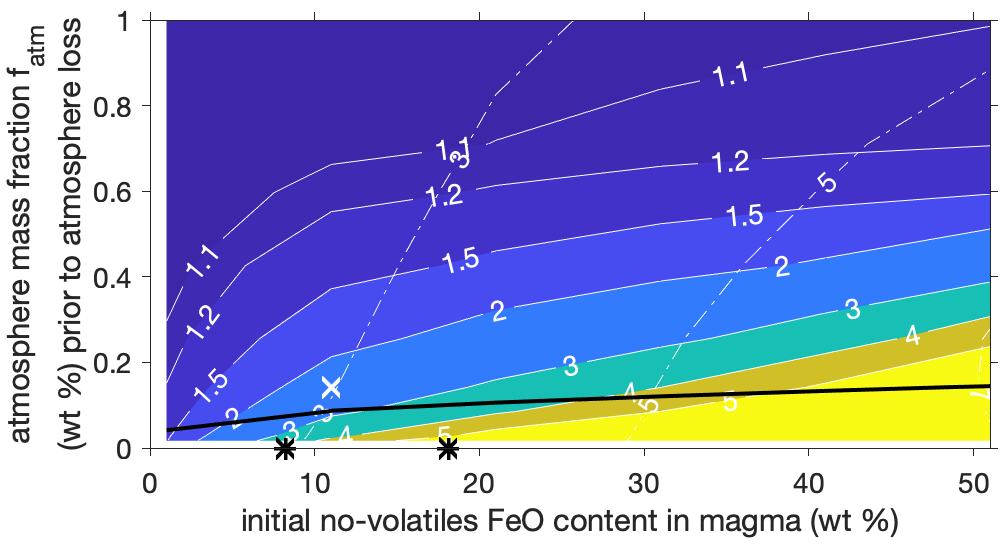}
b)
\includegraphics[trim={0mm 0mm 0mm 0mm},clip,width=1.0\columnwidth]{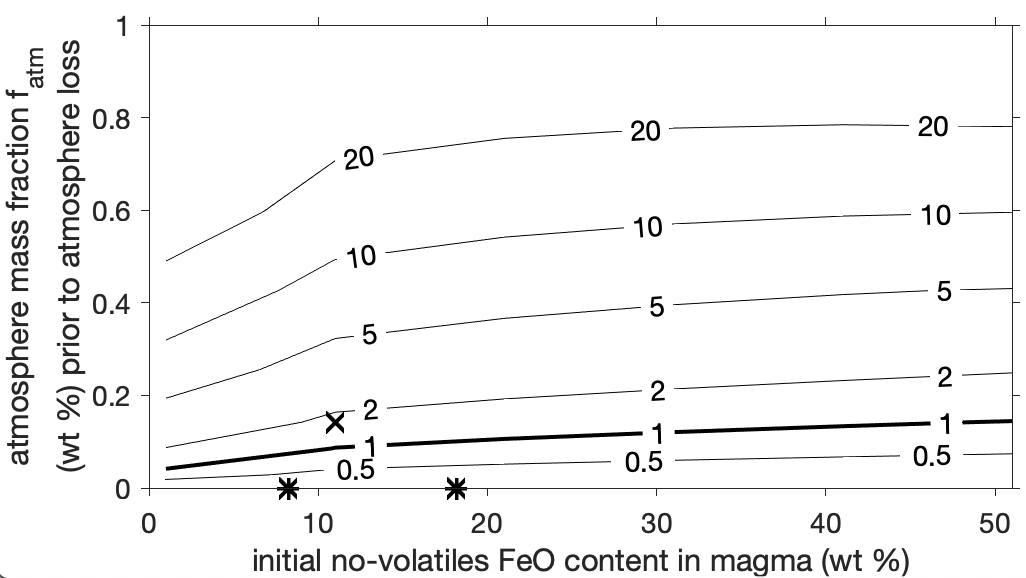}
c)
\includegraphics[trim={0mm 0mm 0mm 0mm},clip,width=1.0\columnwidth]{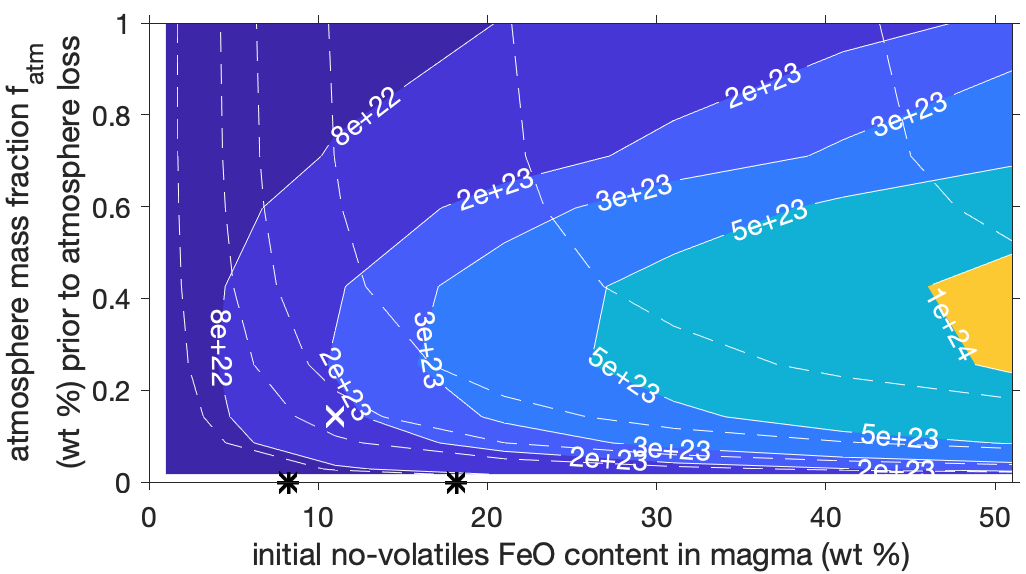}
\caption{Results for 5 $M_\Earth$ simulations. \textbf{(a)}~Colors and solid contours correspond to the ratio, $\chi$, of the mass removal (kg) necessary to dry out an endogenous water-world (combined width of stages ``i'' and ``ii'' on Fig.~\ref{fig:fig1}), to the mass removal needed to remove the H$_2$ envelope from the planet and raise $\mu_{atm}$ above 10 (width of stage ``i'' on Fig.~\ref{fig:fig1}). Dashed contours correspond to $\mu_{atm}$ before any atmospheric loss has occurred. Cross corresponds to the Fig.~\ref{fig:fig1} simulation. The black line is the stoichiometric threshold below which enough FeO is available to destroy all initial H$_2$O. Asterisks on the x-axis correspond to Earth (left) and Mars (right). \textbf{(b)}~Moles of initial H$_2$O on the planet, normalized to the stoichiometric maximum (ignoring thermodynamics) H$_2$O that can be destroyed by reaction with FeO (reaction \ref{eqn:oxidation}). \textbf{(c)}~Mass (kg) of H$_2$O that must be removed to dry out a wet super-Earth (width of stage ``ii'' on Fig.~\ref{fig:fig1}). Dashed contours correspond to the same contours, but for initial H$_2$O mass (i.e., H$_2$O mass at the leftmost edge of Fig.~\ref{fig:fig1}). Overall, for a wide range of parameters, a large initial mass of H$_2$O ensures a relatively-lengthy stage with a 10-2000~bar, H$_2$O-dominated atmosphere.}  
\label{fig:fig2}
\end{figure}

Driven by atmospheric loss, the model atmosphere quickly thins from 5~kbar to 200~bar. At this point almost all of the H$_2$ has been removed and the atmosphere is H$_2$O-dominated ($\mu_{atm}$~$>$~10) (Fig.~\ref{fig:fig1}). Thereafter,  despite ongoing atmospheric loss, the rate of atmospheric \emph{pressure} decline greatly decreases (inflection in Fig.~\ref{fig:fig1} top panel, note log scale). This is because H$_2$O outgasses from the magma. The atmosphere stays water-dominated and buffered (in this run) to 10-100~bars (comparable to an Earth ocean mass of H$_2$O) for a long time. The H$_2$O is not derived from solid volatiles (hydrated minerals or ice). Instead, it is produced on the planet - endogenous water. The total mass that must be removed to devolatilize the planet is $\sim$10$\times$ the maximum mass of the atmosphere. (Water-dominated atmospheres up to 2000 bars are possible by this mechanism; Fig.~\ref{fig:fig2}c). Most of this mass is removed while the planet is a super-Earth-sized, but water-enveloped world. 

For a given star, planets very close to the star will evolve all the way to the right on Fig. 1. Planets at large distances from the star will remain as H$_2$-shrouded worlds with large radii and low densities \citep{Rogers2015}. Planets at intermediate distances from the star will be water-enveloped super-Earths.

A relatively-lengthy stage with a 10-2000~bar, H$_2$O-dominated atmosphere occurs for a wide range of parameters (Fig.~\ref{fig:fig2}a). Only for model runs with $f_{atm}~\gtrsim$~0.6~wt\% is the water-enveloped stage brief. According to models, $f_{atm}~\gtrsim$~0.6~wt\% worlds often stay as sub-Neptunes and do not become super-Earths \citep{RogersOwen2020}. Earth-sized and larger planets should have an initial no-volatiles mantle FeO content of at least a few wt\%. This is because Si from the mantle dissolves into the liquid-Fe core more readily than does mantle O, creating an excess of mantle O that pairs with Fe \citep{Wordsworth2018}. Indeed, Earth, Venus and Mars have FeO mole fraction $\sim$0.1-0.2, and constraints on rocky extrasolar material from contaminated-white-dwarf data give FeO mole fraction 0.1-0.5 (typically 0.15-0.35) \citep{Doyle2020}. Mantle FeO contents of a few wt\% or higher favor a water-enveloped stage in our model (Fig.~\ref{fig:fig2}). 

The typical atmospheric mass during the waterworld stage of a 5~$M_\Earth$ simulation varies between simulations from \mbox{1-16}~$\times$~10$^{21}$~kg (corresponding to 100-2000 bars). The amount of H$_2$O that must be removed to desiccate the waterworld after reaching the high-$\mu_{atm}$ stage  is much larger: 1-3~wt\% of planet mass (2-10~$\times$~10$^{23}$~kg) (Fig.~\ref{fig:fig2}c). This is because H$_2$O is very soluble in magma. 

\begin{figure}
\includegraphics[width=1.02\columnwidth,clip=true,trim={0 0 0 0}]{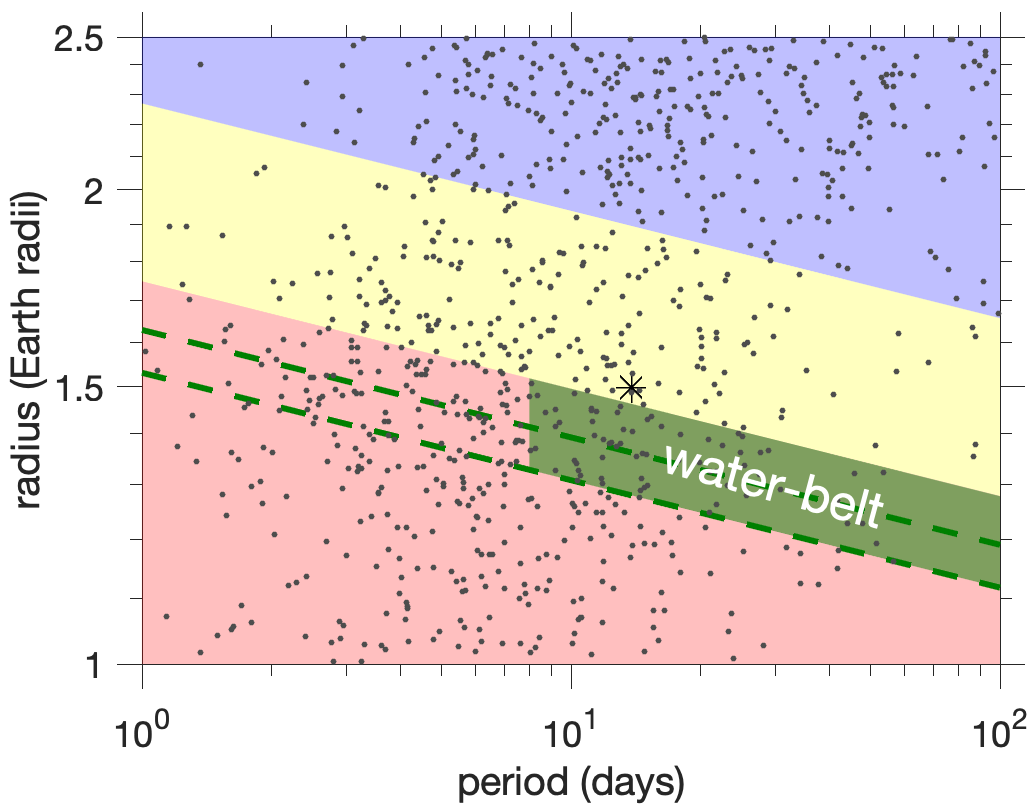}
\caption{Predicted locations of endogenous-water worlds (green) in period-radius space. Planet data from  \citet{FultonPetigura2018} (the radius valley is in reality emptier than shown here, for $<$3-Gyr-old stars). Black asterisk corresponds to Kepler-36b. Radius valley location (yellow) is from \citet{David2020}.  Red indicates super-Earth zone and blue indicates sub-Neptune zone. Predicted endogenous-water worlds define a ``water-belt'' lining the exoplanet radius valley. An initial $f_{atm}$ of 0.43 wt\% is assumed. The belt is wider for an initial no-volatiles FeO content in magma of 30 wt\% (lower dashed line) than for 10 wt\% (upper dashed line).}  
\label{fig:histogram}
\end{figure}

\section{Discussion.}

\subsection{Which hot rocky exoplanets should be water-enveloped today?}

\noindent Endogenous-waterworld predictions can be made in planet radius-period space (Fig.~\ref{fig:histogram}). In radius-period space, rocky exoplanets and sub-Neptunes are separated by a valley that is deficient in planets (\citealt{vanEylen2018}, \citealt{Petigura2020}, \citealt{David2020}). Atmospheric loss drives worlds vertically downward on Fig.~\ref{fig:histogram}. Initially H$_2$-shrouded worlds traverse the valley relatively rapidly as H$_2$ is lost (stage ``i'' in Fig.~\ref{fig:fig1}). Worlds arrive at their near-final radii with 10-2000~bar H$_2$O-dominated atmospheres (stage ``ii'' in Fig.~\ref{fig:fig1}). Whether or not the 10-2000~bar H$_2$O-dominated atmosphere is retained depends on the  amount of irradiation and on $M$. A lower-mass core will lose more atmosphere for the same amount of irradiation. This is because for less massive planets, the atmosphere is more weakly bound. Thus for a given period ($\approx$ irradiation), the radius valley corresponds to a threshold in planet mass ($\approx$ planet core mass). 

At orbital period $p$~=~\mbox{32-64}~days, the paucity of super-Earths with radii $R$~=~1.3-1.6~$R_\Earth$ ($\equiv$~\mbox{3-6}~$M_\Earth$) implies that 3-6~$M_\Earth$ worlds are on average hydrogen-enveloped (and thus have radii 2-3~$R_\Earth$) \citep{RogersOwen2020}. 

In terms of our model, this implies that super-Earths in this period range that have radii just below the valley have received only a little bit more irradiation than is needed to remove all H$_2$, and thus are in the endogenic-water world zone (stage~`ii''). The same reasoning applies to $p$ = 8-32 days if we assume that many cores have mass $>$6~$M_\Earth$ ($\equiv$~1.6~$R_\Earth$). At all periods, sufficiently low-mass worlds will be denuded of nebular-derived H (both as H$_2$ and as H$_2$O). If those low-mass worlds nevertheless have steam atmospheres, then the H$_2$O must be exogenic. In summary, if magma-atmosphere interaction is efficient, then endogenic-water worlds should line the exoplanet radius valley as it evolves over time \citep{David2020}.

To make Fig.~\ref{fig:histogram}, we make the additional, auxiliary assumption that the atmosphere loss process is energy-limited XUV-powered escape (e.g. \citealt{RogersOwen2020}). Specifically, 

\begin{equation}
\mathrm{mass \,removed} = \int \frac{ \eta L}{4 a^2 G} \left(\frac{R^3}{M} \right) \,dt \propto R^{-0.7}
\end{equation}

\noindent where $\eta$ is efficiency, $L$ is star XUV luminosity, $a$ is semimajor axis, and $G$ is the gravitational constant. To get the exponent $R^{-0.7}$, we approximate the atmosphere as thin, planet mass as constant, and $R$ $\propto$ $M^{0.27}$. If the atmosphere mass fraction at the sub-Neptune stage (prior to atmosphere loss) is a constant, then the amount of mass that must be removed scales as $R^{(1/0.27)}$. So for a given $a$, the fractional extent to which volatile mass is removed scales as $R^{-(1/0.27)}\,R^{-0.7}$ = $R^{-4.4}$. Then the lower edge of the water-belt (radius of complete dessication, $R_d$) is approximated by ($R_{lv}$~/~$R_{d}$)~= $\chi^{1/4.4}$ where $R_{lv}$ is the radius of the lower edge of the valley, and $\chi$ is the waterworld multiplier factor from the 5 $M_\Earth$ calculations (the quantity corresponding to the solid contours in Fig.~\ref{fig:fig2}a). This is a simplification: $\chi$ varies with $M$, the atmosphere is not always thin, the location of the radius valley may vary with time \citep{David2020}, and other loss processes may be important. Nevertheless, this calculation shows that the number of planets that is predicted to be endogenous waterworlds can be large (Fig.~\ref{fig:histogram}). Moreover, the predicted water-belt includes Super-Earths that (because of their large radius) are among the best targets for observations.

\subsection{The role of selective escape and Fe$^{3+}$}

\noindent Loss of H$_2$ from the planet oxidizes the magma-plus-atmosphere system. Oxidation of Fe$^{2+}$ to Fe$^{3+}$, 

\vspace{-0.1in}
\begin{equation}
\mathrm{Fe^{2+}O_{(l)}} + 0.25\,\mathrm{O_{2(g)}} = \mathrm{Fe^{3+}O_{1.5(l)}   }
\label{eqn:fe2tofe3}
\end{equation}

\noindent combined with the (thermodynamically disfavored) water-breakdown reaction 

\vspace{-0.1in}
\begin{equation}
\mathrm{2\,H_2O_{(g)}  = 2\,H_{2(g)}+O_{2(g)} }
\label{eqn:waterloss}
\end{equation}

\noindent yields the water-destroying reaction,

\vspace{-0.1in}
\begin{equation}
2\,\mathrm{Fe^{2+}O_{(l)}} + \mathrm{H_2O_{(g)}} = 2\,\mathrm{Fe^{3+}O_{1.5(l)}} + \mathrm{H_{2(g)}}
\label{eqn:oxidation}
\end{equation}

Reaction~\ref{eqn:oxidation} can be important for secondary atmospheres \citep{Schaefer2016,Wordsworth2018}. However, two effects limit its importance for sub-Neptune-to-super-Earth conversion. First, worlds with initial (post-reaction, pre-loss) atmosphere mass fractions $f_{atm}$~$>$0.1~wt\% do not have enough moles of FeO to convert most of the H$_2$O back to H$_2$ (Fig.~\ref{fig:fig2}). Moreover, 0.1~wt\% is close to the minimum $f_{atm}$ for a super-Earth that is imbedded in the nebula \citep{Ginzburg2016}. Therefore it is possible that for real planets FeO oxidation is not enough to destroy much H$_2$O. Second, the ratio of products to reactants for reaction~\ref{eqn:oxidation} is given by the equilibrium constant $K_3$

\vspace{-0.05in}
\begin{equation}
K_3 = \frac{a\mathrm{FeO_{1.5}}^2 \, f_{\mathrm{H2}}}{a\mathrm{FeO}^2 \, f_{\mathrm{H2O}}}
\label{eqn:K}
\end{equation}

\noindent where $a$ denotes chemical activity (effective concentration) and $f$ denotes fugacity (bars). For 1500~K~$<$~$T$~$<$~4000~K, using standard thermodynamics data supplemented by data from \citet{LangeCarmichael1987} and \citet{Zhang2017}, we obtain $K_3$ $\sim$~10$^{-3}$. Thus, if the atmospheric H$_2$/H$_2$O ratio rises above 10$^{-3}$, then eqn.~\ref{eqn:K} implies that the mantle Fe must be mostly FeO (and not FeO$_{1.5}$). Thus reaction~\ref{eqn:oxidation} self-limits. 

Water destruction by reaction~\ref{eqn:oxidation} is most effective if the atmospheric H$_2$/H$_2$O ratio is driven below 10$^{-3}$ by \emph{selective} H escape. Selective H escape at the diffusion limit (assuming a homopause composition of \sfrac{2}{3} H and \sfrac{1}{3} O) allows water destruction at a rate of $\sim$3~$\times$~10$^{22}$~kg~Gyr$^{-1}$ \citep{Wordsworth2018}. This limit permits enough H escape to overcome the Fe$^{2+}$ sink for many worlds below the thick black line in Fig.~\ref{fig:fig2}b. Whether or not this upper limit is approached over the lifetime of the planet will depend on the XUV flux, among other factors \citep{Wordsworth2018}. If all H$_2$O is destroyed, the end state is a solid surface rich in the  Fe$^{3+}$ minerals, such as magnetite and hematite. 

Worlds above the thick black line in Fig.~\ref{fig:fig2} do not have enough FeO to reduce all the water. For these worlds, if~Fe$^{2+}$ is completely oxidized to Fe$^{3+}$, then further selective H escape leaves behind an atmosphere enriched in O$_2$ (e.g., \citealt{Wordsworth2018}).

\vspace{-0.05in}
\subsection{Uncertainties and Limitations}
\noindent Some magma-atmosphere interaction is unavoidable. The magma is almost inviscid and as the magma-atmosphere interface is depressurized and cooled, bubbles and cold boundary layers form, and these stir up deeper magma. However, we do not know if the magma ocean is well-stirred enough that each magma parcel equilibrates with the atmosphere at the maximum atmospheric pressure. (A sensitivity test reducing the volume of magma that interacts with the atmosphere three-fold causes a roughly two-fold reduction in the relative length of the water-dominated stage). An alternative endmember hypothesis is that  sub-Neptunes cores are stratified \citep{Ormel2020}.  

An intermediate possibility is that magma-atmosphere interaction occurs Gyr after planet formation. Magma-atmosphere interaction can cause factor-of-several changes in $f_{atm}$ even without atmospheric loss \citep{Kite2020}. Perhaps this contributes to the formation and evolution of the radius valley. Delaying magma-atmosphere interaction until most of the atmosphere has already been lost drives planets down on the plots in Fig.~\ref{fig:fig2}, making magnetite/hematite surfaces more likely. These possibilities motivate more research on mixing within magma oceans on sub-Neptunes. 

As the atmosphere shifts from H$_2$-dominated to H$_2$O dominated, the efficiency with which loss drivers are converted into mass loss might decrease \citep{Johnstone2018,Johnstone2020}. This does not affect our conclusions.

We approximate magma-atmosphere interface temperature and within-envelope gravitational acceleration as constant during atmosphere loss (2500~K, and 1.2$^{-2}$ of the interface value, respectively). We neglect changes in the fugacity coefficient during atmosphere loss \citep{Kite2019}. For this study, these effects are small relative to that of H$_2$O's greater solubility.

We ignore magma ocean solidification, which pushes volatiles into the atmosphere as the magma surface cools \citep{Turbet2020}. 

The starting point for our calculations is the results of \citet{Kite2020}. The most important limitation of that study is that material property data (e.g. solubilities) are based on a relatively small number of laboratory and/or numerical experiments.

Recent work indicates magma Fe$^{2+}$ disproportionates at high pressure \citep{Armstrong2019}. This process leads to a more-oxidized magma and more-oxidized atmosphere, supporting our conclusion \citep{Armstrong2019}.

We ignore gases other than H$_2$ and H$_2$O. This is justified because C species are sparingly soluble in magma \citep{KepplerGolabek2019}, so they are not protected by dissolution in the magma. Instead they are lost with the H$_2$.

\vspace{-0.02in}
\subsection{Tests and Implications}
\noindent We describe three possible tests of our model's predictions (Fig.~\ref{fig:histogram}). 

(1) Whereas hot rocky exoplanets with solid-derived volatiles should have volatile C/O $\sim$ 1 \citep{Bitsch2019}, endogenous-water worlds should have lower volatile C/O ratio, with atmospheres that are $>$50\% H$_2$O by number. This hypothesis can be tested using spectroscopic observations (e.g., \citealt{Benneke2019}).

(2) The 10-2000 bar atmospheres predicted by our model are 150-500 km thick (according to \citealt{Turbet2020}), enough to make a 1.5 $R_\Earth$ planet less dense relative to an Earth-like composition by up to 17\%. A statistical test of this prediction is imminently possible. Already, radial-velocity data for transiting super-Earths \citep{Sinukoff2018, Otegi2020} show a clump with densities too low for Earthlike composition. These densities imply an atmosphere, because although they are consistent with spheres of pure MgSiO$_3$, cosmochemically plausible bare-rock planets contain Fe and so are denser. Whether this clump in the data is real or not awaits a detailed analysis correcting for detection biases. Our model predicts that more massive rocky planets will show this density anomaly more frequently.

(3) Tests for predicted atmospheres can be done using phase curves or measurements of secondary-eclipse depth (e.g., \citealt{Kreidberg2019,Koll2019, Mansfield2019}). These methods do not require detection of atmospheric molecules. A survey of $\sim$10 rocky exoplanets within the water-belt in Fig.~\ref{fig:histogram} would be sufficient to test the prediction that most have atmospheres.

If endogenous-high-$\mu_{atm}$ worlds are common, then implications include the following. (1)~Water envelopes can be made in situ, so water is not a proxy for migration of solid-derived volatiles across the snowline. (2)~Many bare-rock worlds will have surfaces rich in the Fe$^{3+}$-bearing, spectroscopically-distinctive minerals magnetite and hematite \citep{Mansfield2019}. (3)~Since endogenous-water worlds imply magma-atmosphere interaction, the magma-atmosphere interaction explanation for the sharp drop-off in exoplanet abundance above 3~$R_\Earth$ \citep{Kite2019} would be favored. (4)~Endogenous-water formation should work as well or better in the habitable zone. Therefore, habitable-zone oceans that have H$_2$O with nebula-sourced H will be common \citep{IkomaGenda2006}. 
 
 \vspace{-0.01in}

\begin{figure}
 \includegraphics[width=1.05\columnwidth]{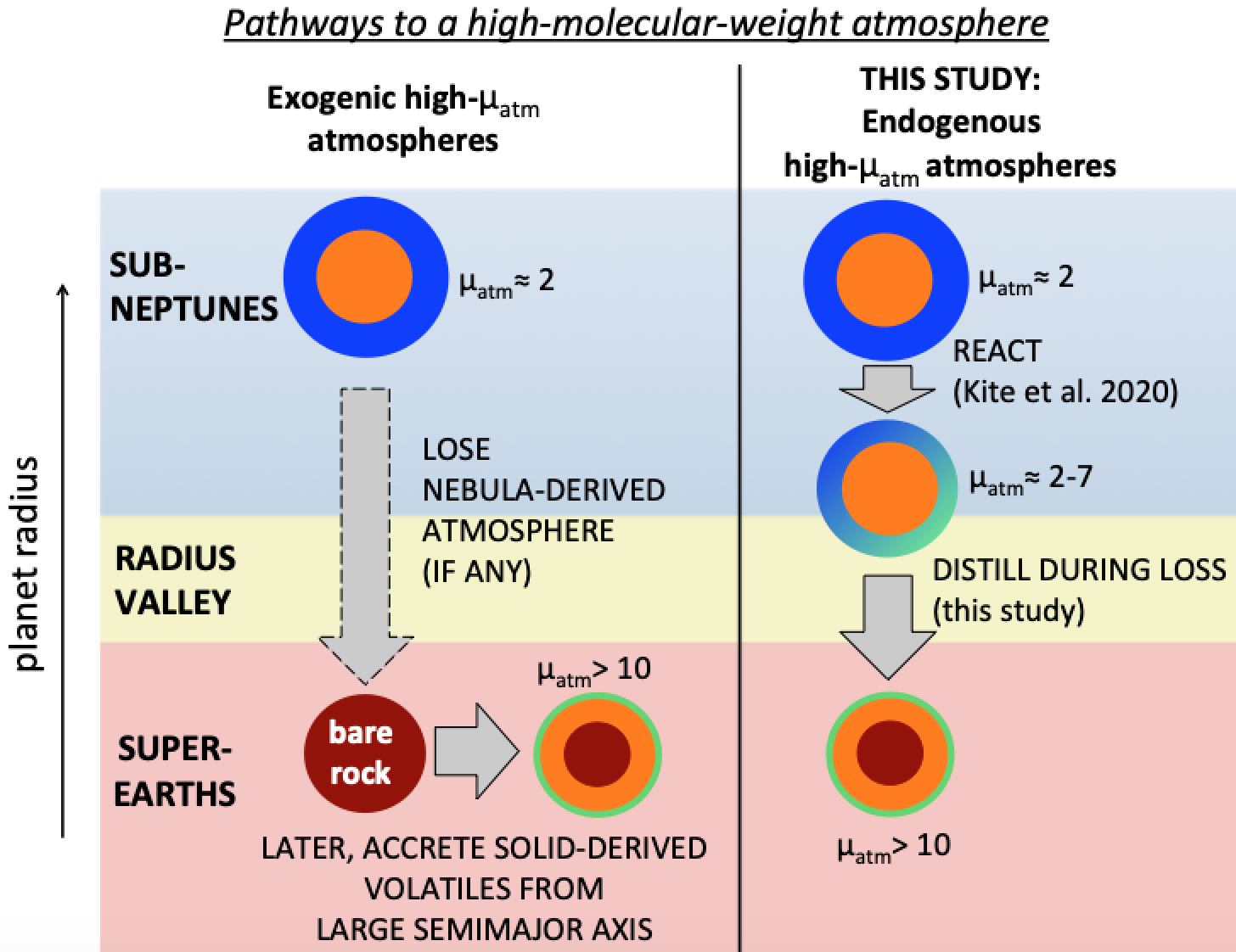}
\caption{Graphical abstract of this paper. Blue and green shading corresponds to low molecular weight species and high molecular weight species, respectively. Our proposed scenario provides an alternative route to producing high-mean-molecular-weight atmospheres on hot rocky exoplanets. The hot super-Earth population could contain both types of atmospheres.}
\label{fig:cartoonsummary}
\end{figure}

\section{Summary and Conclusion.}

\noindent 
In our model, adding and then removing a H$_2$ atmosphere draws out a H$_2$O atmosphere. This outcome can be understood as follows: (1)~Early magma-atmosphere reaction transfers electrons from H$_2$ to Fe$^{2+}$, forming dense Fe$^0$ metal which sinks to the liquid-iron core. This oxidation of the magma-atmosphere system is irreversible because (by assumption) chemical re-equilibration between the liquid-iron core and the magma is forbidden. (2)~The H$_2$O atmosphere is buffered by dissolved H$_2$O. The dissolved reservoir is massive because H$_2$O is soluble in magma and because the FeO-H$_2$ reaction volatilizes the oxygen from the magma, which increases the mass of volatiles. Our result is different from that of \citet{KiteBarnett2020}, which traced the fate of a hypothetical 10$^{21}$~kg dose of solid-derived volatiles, simply because the mass of H$_2$O is so much larger. 
 
In conclusion, if magma-atmosphere interaction is efficient early in a sub-Neptune's history, then during sub-Neptune-to-super-Earth conversion $>$80\% of the reduction in atmospheric volume occurs while most of the volatile mass remains. Thus, location below the radius valley does not imply bare-rock status. Instead, super-Earths can maintain 150-500~km thick (10-2000~bar) H$_2$O atmospheres, buffered by a large reservoir of dissolved-in-magma H$_2$O. The H$_2$O is assembled from nebula-derived H and magma-sourced O. The main weakness of the idea is that we do not know if magma-atmosphere interaction is efficient on small sub-Neptunes. The main strength of the endogenous-water-world hypothesis is that it is imminently testable.

\vspace{0.1in}
\noindent \emph{Acknowledgements.} 
We thank B. Fegley and E. Ford. We thank the anonymous reviewer for comments that substantially improved the presentation of the results. Grants: NASA (NNX16AB44G). 
\vspace{0.05in}

\noindent \emph{Code availability.} All code can be obtained for unrestricted further use by emailing E.S.K.

\renewcommand\thefigure{A\arabic{figure}}    
\setcounter{figure}{0}

\clearpage

\end{document}